\documentclass[pdftex, aps, prd, letterpaper, superscriptaddress, nofootinbib, twocolumn,
               preprintnumbers, longbibliography]{revtex4-1}
\pdfoutput=1
\usepackage{amsmath,amssymb,amsbsy}
\usepackage[utf8]{inputenc}
\usepackage{graphicx,booktabs,longtable,tabularx}
\usepackage{soul,xcolor,topcapt}
\usepackage[colorlinks=true, linkcolor=blue, filecolor=blue, urlcolor=blue, citecolor=blue, pdftex, plainpages=false]{hyperref}

\newcommand{\svev}[1]{\left\langle #1\right\rangle}

\definecolor{orange}{rgb}{1.0, 0.5, 0}

\begin{document}
\preprint{FERMILAB-PUB-24-0098-V}

\title{Infrared fixed point in the massless twelve-flavor SU(3) gauge-fermion system}
\author{Anna Hasenfratz}
\email{anna.hasenfratz@colorado.edu}
\affiliation{Department of Physics, University of Colorado, Boulder, Colorado 80309, USA}
\author{Curtis T.~Peterson}
\email{curtis.peterson@colorado.edu}
\affiliation{Department of Physics, University of Colorado, Boulder, Colorado 80309, USA}

\begin{abstract}
We present strong numerical evidence for the existence of an infrared fixed point in the renormalization group flow of the SU(3) gauge-fermion system with twelve massless fermions in the fundamental representation. Our numerical simulations using nHYP-smeared staggered fermions with Pauli-Villars improvement do not exhibit any first-order bulk phase transition in the investigated parameter region. We utilize an infinite volume renormalization scheme based on the gradient flow transformation to determine the renormalization group $\beta$ function. We identify an infrared fixed point at $g^2_{\mathrm{GF}\star}=6.60(62)$ in the GF scheme and calculate the leading irrelevant critical exponent $\gamma_{g}^{\star}=0.199(32)$. Our prediction for $\gamma_{g}^{\star}$ is consistent with available literature at the $1\mbox{-}2\sigma$ level.
\end{abstract}

\maketitle

\section{Introduction}

The infrared properties of the SU(3) gauge-fermion system with $N_f=12$ massless fundamental flavors has been studied extensively using a variety of analytical and numerical techniques. Such techniques include perturbation theory \cite{Ryttov:2010iz, Pica:2010xq, Ryttov:2016ner, Ryttov:2017kmx, Ryttov:2016hal, DiPietro:2020jne}, use of the gap equation \cite{Appelquist:1998rb, Bashir:2013zha}, functional renormalization group methods \cite{Braun:2006jd, Braun:2010qs}, conformal expansion \cite{Lee:2020ihn}, conformal bootstrap \cite{Li:2020bnb}, the background field method \cite{Grable:2023eyj}, perturbative non-relativistic quantum chromodynamics \cite{Chung:2023mgr}, large-$N$ expansion \cite{Romatschke:2024yhx}, and nonperturbative lattice simulations \cite{Appelquist:2007hu, Appelquist:2009ty, Lin:2012iw, Lin:2015zpa, Fodor:2016zil, Hasenfratz:2016dou, Fodor:2017nlp, Fodor:2017gtj, Hasenfratz:2017qyr, Hasenfratz:2019dpr, Hasenfratz:2010fi, Hasenfratz:2011xn, Hasenfratz:2019puu, Deuzeman:2009mh, Fodor:2011tu, Appelquist:2011dp, DeGrand:2011cu, Fodor:2012et, Aoki:2012eq, LatKMI:2013bhp, Cheng:2013xha, Lombardo:2014pda, Fodor:2009wk, Cheng:2013eu}. Investigations based on lattice simulations have utilized finite-volume step-scaling \cite{Appelquist:2007hu, Appelquist:2009ty, Lin:2012iw, Lin:2015zpa, Fodor:2016zil, Hasenfratz:2016dou, Fodor:2017nlp, Fodor:2017gtj, Hasenfratz:2017qyr, Hasenfratz:2019dpr}, Monte Carlo renormalization group methods \cite{Hasenfratz:2010fi, Hasenfratz:2011xn}, hadron mass and decay constant spectroscopy \cite{Deuzeman:2009mh, Fodor:2011tu, Appelquist:2011dp, DeGrand:2011cu, Fodor:2012et, Aoki:2012eq, LatKMI:2013bhp, Cheng:2013xha, Lombardo:2014pda}, and the Dirac eigenmode spectrum \cite{Fodor:2009wk, Cheng:2013eu}. Many investigations suggest that the $N_f=12$ system is infrared conformal\footnote{See Refs. \cite{Ryttov:2010iz, Pica:2010xq, Ryttov:2016ner, Ryttov:2017kmx, DiPietro:2020jne, Appelquist:1998rb, Bashir:2013zha, Lee:2020ihn, Li:2020bnb, Appelquist:2007hu, Appelquist:2009ty, Lin:2012iw, Hasenfratz:2016dou, Hasenfratz:2017qyr, Hasenfratz:2019dpr, Hasenfratz:2010fi, Hasenfratz:2011xn, Deuzeman:2009mh, Appelquist:2011dp, DeGrand:2011cu, Aoki:2012eq, LatKMI:2013bhp, Cheng:2013xha, Lombardo:2014pda, Cheng:2013eu}}, though a minority of studies conclude that the system is confining with chiral symmetry breaking, or are inconclusive, as they find neither direct evidence of chiral symmetry breaking nor of an infrared fixed point\footnote{See Refs. \cite{Grable:2023eyj, Romatschke:2024yhx, Fodor:2011tu, Lin:2015zpa, Fodor:2012et, Fodor:2016zil, Fodor:2019ypi}}. 

Most lattice studies are affected by the presence of a bulk first-order phase transition \cite{Deuzeman:2012ee, NunesdaSilva:2012mkt, Schaich:2012fr, Rindlisbacher:2021skt, Rindlisbacher:2023qjg, Springer:2023hcc}. Such unphysical phase transitions are triggered by strong ultraviolet fluctuations in the fermion sector that prevent lattice simulations from reaching deep into the infrared regime. Even when strong couplings are reached, lattice cutoff effects make it difficult to take the proper continuum limit, leading to inconsistent results between different lattice formulations. It is imperative to reduce the ultraviolet fluctuations that trigger first-order bulk phase transitions. Ref. \cite{Hasenfratz:2021zsl} suggested to include unphysical heavy Pauli-Villars fields to achieve the necessary improvement. Lattice Pauli-Villars fields are similar to their continuum analogue - they have the same action as the fermions, but possess bosonic statistics. Their mass is at the level of the cutoff. Therefore, they decouple in the infrared limit, while in the ultraviolet they compensate for cutoff effects introduced by the fermions. This idea has been tested in simulations of the SU(3) gauge-fermion system with ten massless fundamental Dirac fermions (flavors) and the SU(4) gauge-fermion system four massless fundamental and four massless two-index Dirac fermions \cite{Hasenfratz:2023sqa, Hasenfratz:2023wbr}. Both studies extended the reach into the infrared regime significantly,  and both present clear evidence for infrared conformality in those systems. See also Refs. \cite{Rindlisbacher:2021skt, Rindlisbacher:2023qjg} for an alternative proposal to remove unphysical bulk phase transitions. 

\begin{figure}
    \centering
    \includegraphics[width=\columnwidth]{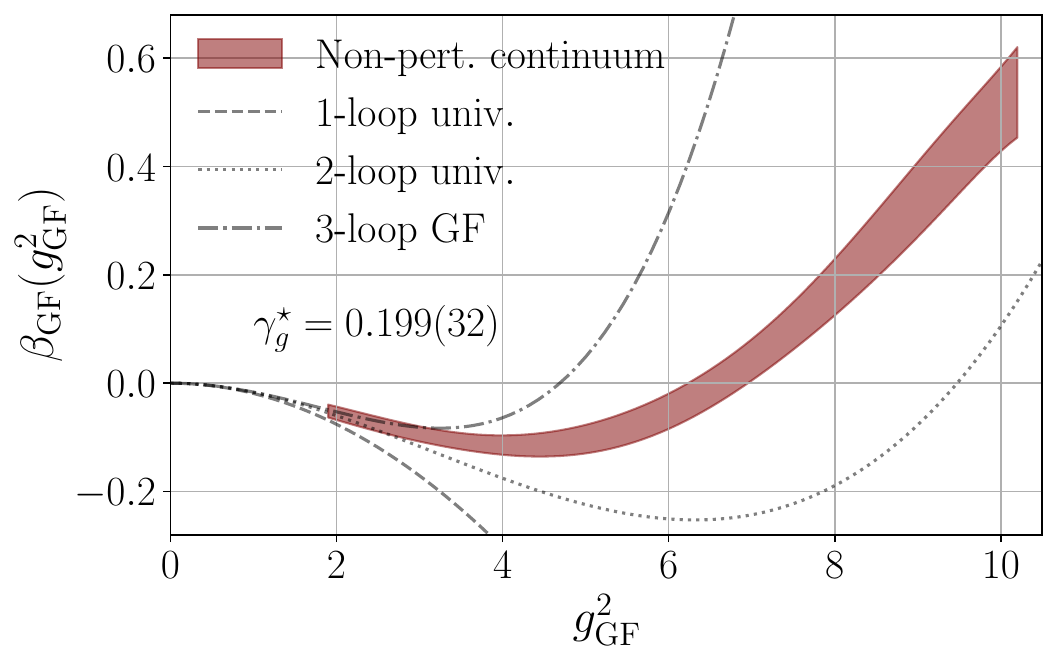}
    \caption{Our continuum prediction for $\beta_{\mathrm{GF}}\big(g^2_{\mathrm{GF}}\big)$ as a function of $g^2_{\mathrm{GF}}$ (maroon band) juxtaposed against the 1- (dashed), 2- (dotted) and 3-loop (dashed-dotted) gradient flow $\beta$ function from perturbation theory \cite{Harlander:2016vzb}. The width of the maroon band indicates the error. Also given is our prediction for the leading irrelevant critical exponent at the IRFP, $\gamma_{g}^{\star}=0.199(32)$. }
    \label{fig:final_result}
\end{figure}

In this work, we utilize Pauli-Villars (PV) improvement to study the infrared properties of the massless SU(3) gauge-fermion system with $N_f=12$ fundamental flavors. We calculate the renormalization group (RG) $\beta$ function, defined as the logarithmic derivative of the renormalized running coupling $g^2(\mu)$ with respect to an energy scale $\mu$ as
\begin{equation}
    \beta\big(g^2\big) = \mu^2\frac{\mathrm{d}g^2(\mu)}{\mathrm{d}\mu^2}.
\end{equation}
If the $\beta$ function exhibits a fixed point at some coupling $g^2_{\star}$, the gauge coupling becomes irrelevant and the system is infrared conformal.
The value of $g^2_{\star}$ depends on the renormalization scheme, but the existence of the infrared fixed point (IRFP) does not.

On the lattice, it is convenient to use gradient flow (GF) transformation \cite{Narayanan:2006rf, Luscher:2009eq, Luscher:2010iy} as a continuous smearing operation to define an infinite-volume renormalization scheme \cite{Fodor:2017die, Carosso:2018bmz, Carosso:2019qpb, Hasenfratz:2019puu, Hasenfratz:2019hpg, Hasenfratz:2022wll}. In the GF scheme, the renormalized coupling $g^2_{\mathrm{GF}}(t)$ in infinite volume at flow time $t \propto 1/\mu^2$ is defined in terms of the Yang-Mills energy density $E(t)$ as
\begin{equation}
    g^2_{\mathrm{GF}}(t) \equiv \mathcal{N}\langle t^2E(t) \rangle,
\end{equation}
where $\mathcal{N}=128\pi^2/(3N_c^2-3)$ is a constant chosen to match $g^2_{\mathrm{GF}}$ to $g^2_{\overline{\mathrm{MS}}}$ at tree-level \cite{Luscher:2009eq}. The corresponding RG $\beta$ function is
\begin{equation}
    \beta_{\mathrm{GF}}\big(g_{\mathrm{GF}}^2\big) \equiv -t\frac{\mathrm{d}g_{\mathrm{GF}}^2(t)}{\mathrm{d}t}. 
\end{equation}
To calculate the infinite volume gradient flow $\beta$ function $\beta_{\mathrm{GF}}\big(g_{\mathrm{GF}}^2\big)$ from finite-volume simulations, we utilize the continuous $\beta$ function method (CBFM) proposed in Refs. \cite{Fodor:2017die, Hasenfratz:2019puu, Hasenfratz:2019hpg} and deployed extensively in Refs. \cite{Kuti:2022ldb, Hasenfratz:2023bok, Wong:2023jvr, Hasenfratz:2023sqa, Hasenfratz:2023wbr, Peterson:2021lvb} to a variety of strongly-coupled gauge-fermion systems. In this paper we follow the steps described in \cite{Peterson:2021lvb, Hasenfratz:2023bok} with additional extensions for improved error estimation. We discuss the continuous $\beta$ function method and its implementation in further detail in Sec. \ref{sec:beta_function}. 

As a preview, we show our nonperturbative prediction for $\beta_{\mathrm{GF}}\big(g^2_{\mathrm{GF}}\big)$ as a function of $g^2_{\mathrm{GF}}$ in Fig. \ref{fig:final_result}. The predicted $\beta$ function converges to the universal 1-/2-loop and 3-loop gradient flow perturbative $\beta$ functions  at small $g^2_{\mathrm{GF}}$ \cite{Harlander:2016vzb}. Around $g^2_{\mathrm{GF}\star}=6.60(62)$, the nonperturbative $\beta$ function unambiguously exhibits an infrared fixed point. From the slope of $\beta_{\mathrm{GF}}\big(g^2_{\mathrm{GF}}\big)$ at $g^2_{\mathrm{GF}\star}$, we calculate the leading irrelevant critical exponent $\gamma_{g}^{\star}=0.199(32)$ and find that it is consistent with the perturbative calculations of Refs. \cite{Ryttov:2017kmx, DiPietro:2020jne} at the $1\sigma$ level and the lattice calculation of Ref. \cite{Hasenfratz:2016dou} at the $2\sigma$ level. We control for systematic errors in the infinite volume extrapolation step of the CBFM using Bayesian model averaging. Additionally, our Pauli-Villars improved simulations offer tight control over systematics in the continuum extrapolation step of the CBFM.  Our data is publicly available at Ref. \cite{peterson_2024_10719052}.

This paper is laid out as follows. In Sec. \ref{sec:numerical_details}, we summarize details of our numerical simulations. In Sec. \ref{sec:beta_function}, we review the continuous $\beta$ function method and discuss our analysis. We explain our calculation of the leading irrelevant critical exponent in Sec. \ref{sec:leading_irrel_exp}, and wrap up in Sec. \ref{sec:conclusions} with conclusions.

\section{Numerical details}\label{sec:numerical_details}

\begin{table}[tb]
\centering
\begin{tabular}{c@{~~~~~~~~}c@{~~~~~~~~~~}c@{~~~~~~~~~~}c@{~~~~~~~~~~}c@{~~~~~~~~~~}c}
  \hline\hline
$\beta_b$  &\multicolumn{5}{c}{$L/a$} \\
  \cline{2-6}
  & 24     & 28      & 32      & 36      &  40 \\
\hline
9.20 & 340 & 253 & 188 & 188 & 133 \\
9.40 & 347 & 262 & 215 & 273 & 186 \\
9.60 & 244 & 233 & 251 & 203 & 166 \\
9.80 & 275 & 329 & 250 & 297 & 280  \\
10.0 & 271 & 246 & 312 & 151 & 134 \\
10.2 & 184 & 209 & 217 & 221 & 133 \\
10.4 & 283 & 241 & 299 & 221 & 142 \\
10.8 & 246 & 220 & 288 & 208 & 306 \\
11.0 & 236 & 288 & 156 & 151 & 156 \\
11.4 & 188 & 194 & 223 & 193 & 183 \\
12.0 & 182 & 248 & 200 & 254 & 167 \\
12.8 & 180 & 179 & 204 & 254 & 209 \\
13.6 & 251 & 183 & 168 & 254 & 228 \\
14.6 & 253 & 191 & 178 & 251 & 226 \\
\hline\hline
\end{tabular}
\caption{The number of thermalized configurations analyzed at each bare coupling $\beta_b$ and volume $L/a$. The configurations are separated by 10 MDTUs.} \label{table:ensembles}
 \end{table}

\begin{figure}
    \centering
    \includegraphics[width=\columnwidth]{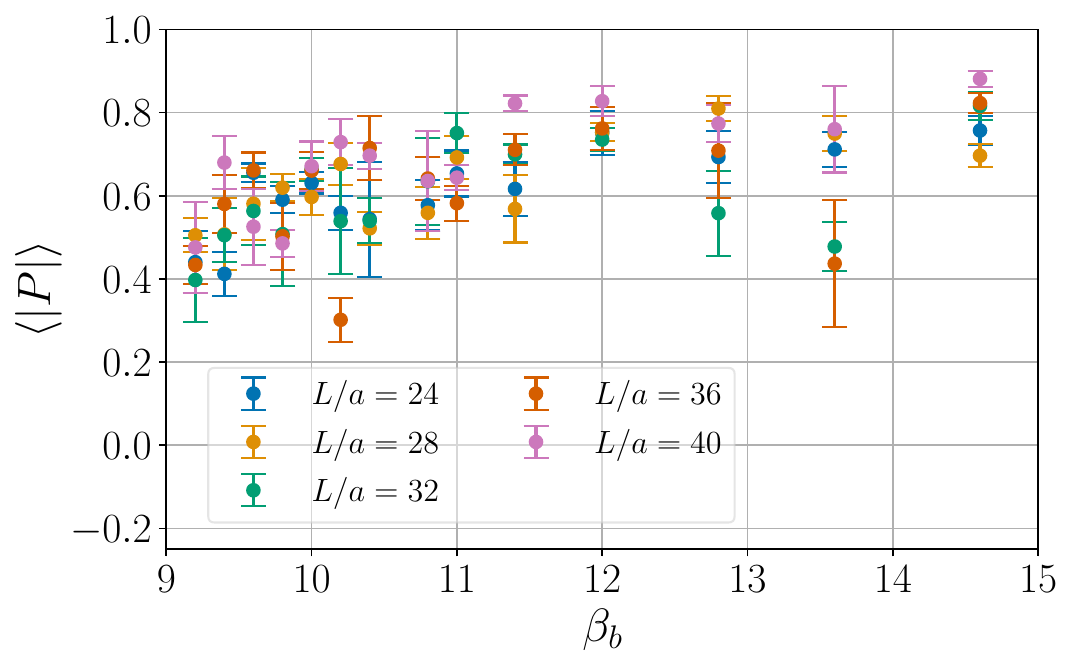}
    \caption{The gradient-flowed Polyakov loop expectation value at flow time $8t/a^2 \approx (L/2a)^2$ versus the bare gauge coupling $\beta_b$ on each volume in Table \ref{table:ensembles}. The absolute value of the Polyakov loop is shown by colored error bars: $L/a=24$ (blue), $28$ (yellow), $32$ (green), $36$ (orange), and $40$ (pink).}
    \label{fig:poly}
\end{figure}

We simulate the massless twelve-flavor SU(3) gauge-fermion system using an adjoint-plaquette gauge action with $\beta_{F}/\beta_{A}=-0.25$ ($\beta_{F} \equiv \beta_b$) and a massless ($am_f=0.0$) nHYP-smeared staggered fermion action with four massive ($am_{\mathrm{PV}}=0.5$) Pauli-Villars (PV) fields per staggered fermion \cite{Hasenfratz:2001hp,Hasenfratz:2007rf,Cheng:2011ic,Hasenfratz:2021zsl}. The ``pions" of these PV fields have mass $a m^{(\mathrm{PV})}_{\mathrm{PS}}\gtrsim 1.0$\footnote{$\mathrm{PS}=\mathrm{pseudoscalar}$} and generate gauge loops in the effective gauge action with a size that decays exponentially with $a m^{(\mathrm{PV})}_{\mathrm{PS}}$ \cite{Hasenfratz:2021zsl}. As long as the volume is much larger than $1/m^{(\mathrm{PV})}_{\mathrm{PS}}$ and $m_f \ll m^{(\mathrm{PV})}_{\mathrm{PS}}$ the PV fields decouple in the infrared. Their only effect is a modified, but local, gauge action. One of the goals of the present work is to illustrate the validity of this expectation.

We use antiperiodic boundary conditions in all four directions for both the staggered fermion fields and PV fields. Our numerical simulations are performed using the hybrid Monte Carlo algorithm \cite{Duane:1987de} implemented in a modified version of the \texttt{MILC} library\footnote{The modified \texttt{MILC} library can be found at \url{https://github.com/daschaich/KS_nHYP_FA}} and the \texttt{Quantum EXpressions} (\texttt{QEX}) library\footnote{Our fork of \texttt{QEX} can be found at \url{https://github.com/ctpeterson/qex}} \cite{Osborn:2017aci}. We set the molecular dynamics trajectory length to $\tau=1.0$. Our configurations are separated by ten trajectories (10 molecular dynamics time units). We perform our simulations at fourteen bare gauge couplings ($9.20 \leq \beta_{b} \leq 14.6$) and five symmetric volumes ($24 \leq L/a \leq 40$). In Table \ref{table:ensembles}, we list the total number of thermalized configurations on each ensemble. 

Our gradient flow measurements are performed using either the modified \texttt{MILC} or \texttt{QEX} libraries \cite{Osborn:2017aci}. We flow our configurations using Wilson flow \cite{Luscher:2009eq, Luscher:2010iy}, integrating the gradient flow equations using the 4th-order Runge-Kutta algorithm discussed in Ref. \cite{Luscher:2009eq} with time step $\mathrm{d}t/a^2=0.02$ for $t/a^2 \leq 5.0$ and $\mathrm{d}t/a^2=0.1$ for $t/a^2>5.0$. At each integration step, we measure the Yang-Mills energy density $E(t)$ using the Wilson (W) and clover (C) discretizations. In the rest of this paper, we refer to results based on Wilson flow and Wilson operator as WW, while we refer to results based on Wilson flow and clover operator as WC. Our data for the Yang-Mills energy density $E(t)$ from the Wilson and clover operator is available at Ref. \cite{peterson_2024_10719052}. 

In Fig. \ref{fig:poly} we plot the expectation value of the magnitude of the gradient-flowed Polyakov loop  $|P|$ at  $8t/a^2 \approx (L/2a)^2$ against the bare gauge coupling $\beta_b$ for all ensembles in Table. \ref{table:ensembles}. Errors are estimated using the ``$\Gamma$-method'' technique implemented by the \texttt{pyerrors} package \cite{Wolff:2003sm, Schaefer:2010hu, Ramos:2020scv, Joswig:2022qfe, Joswig_pyerrors_A_python_2023}; however, they are likely underestimated. Nonetheless, the Polyakov loop suggests that the system is not confining in the range of couplings and volumes that we use in the present study.

\section{Nonperturbative $\beta$ function}\label{sec:beta_function}

\begin{figure*}[tb]
    \centering
    \includegraphics[width=\columnwidth]{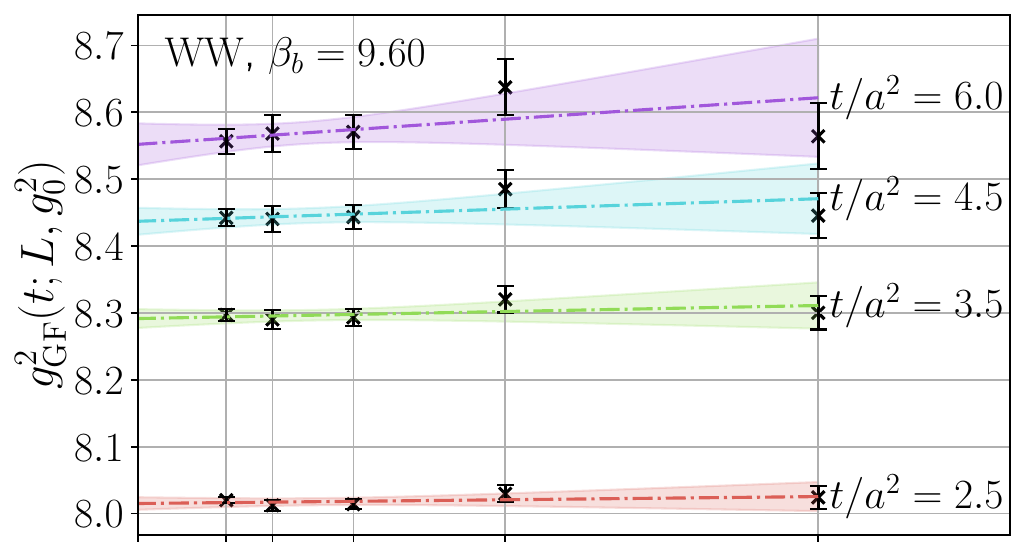}\includegraphics[width=\columnwidth]{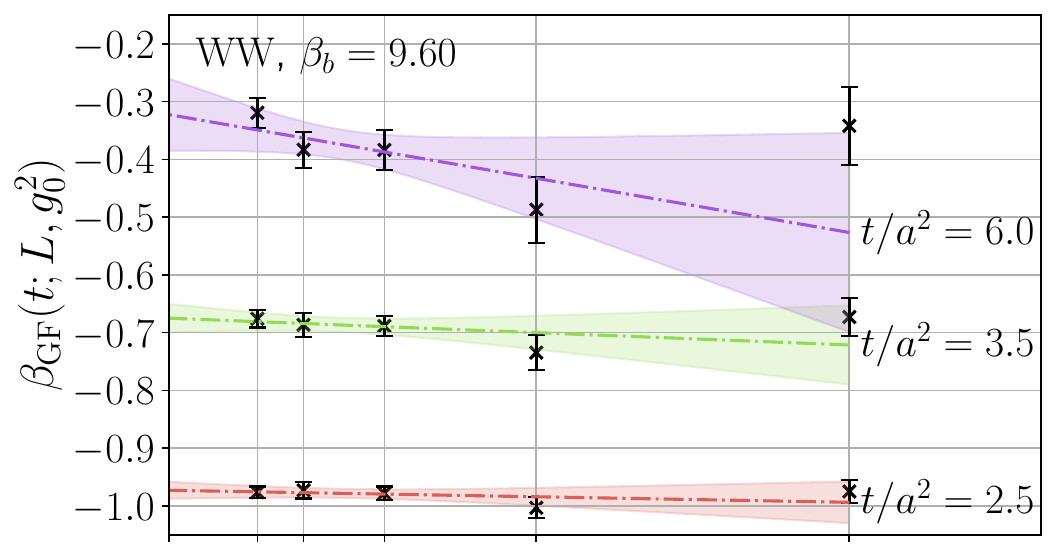}
    \includegraphics[width=\columnwidth]{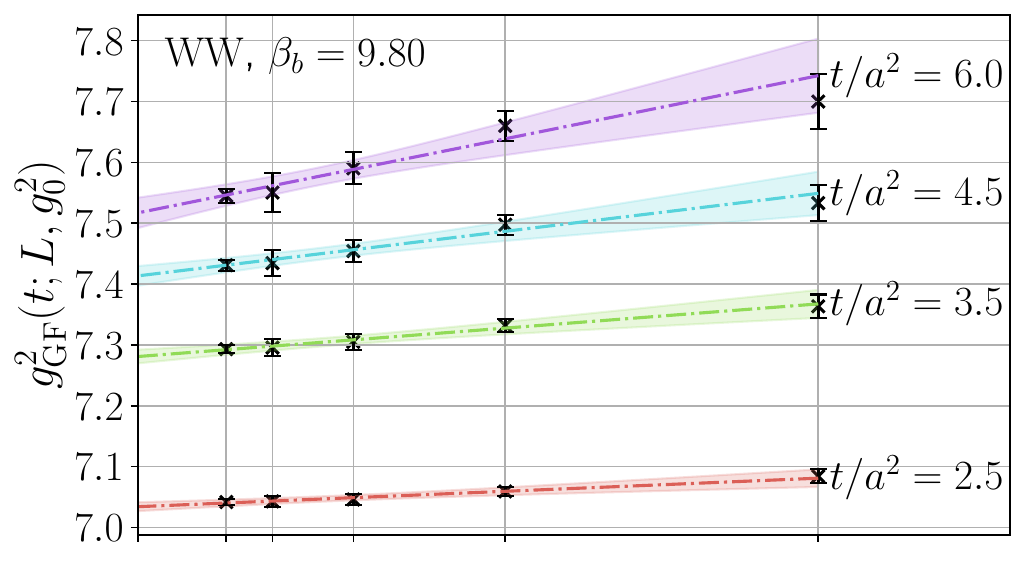}\includegraphics[width=\columnwidth]{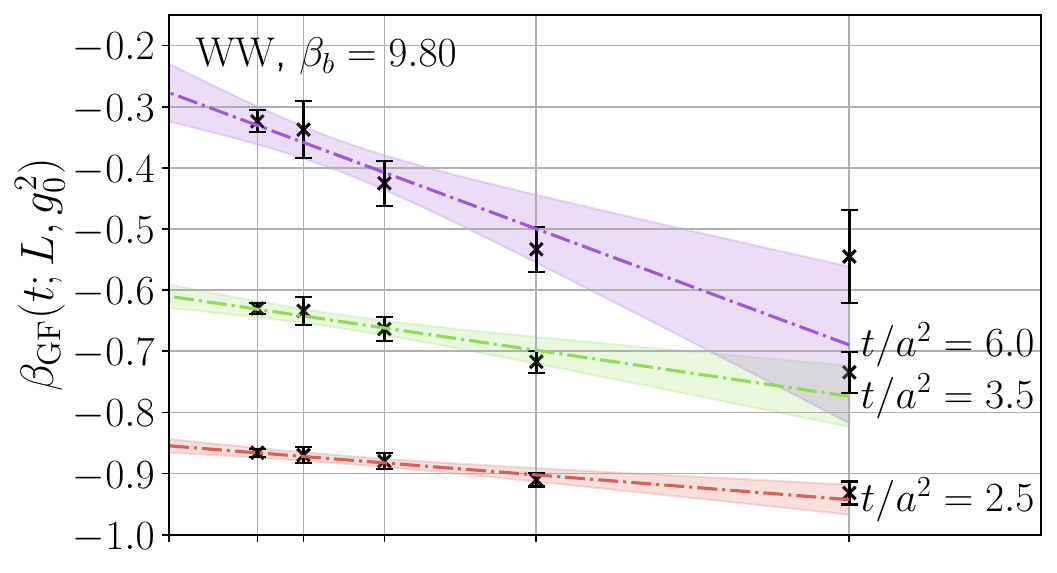}
    \includegraphics[width=\columnwidth]{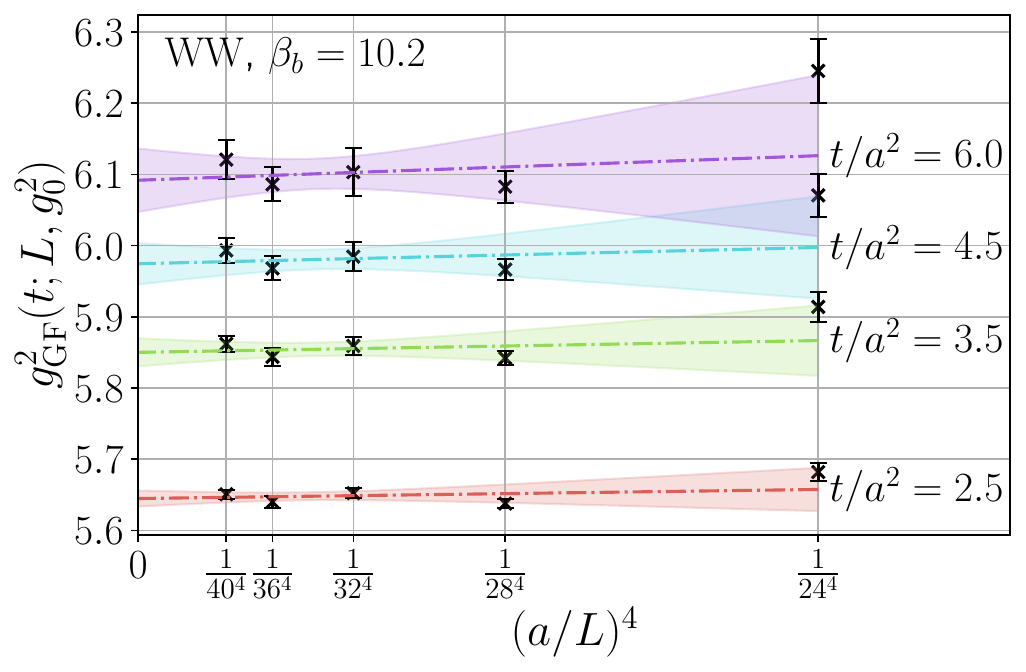}\includegraphics[width=\columnwidth]{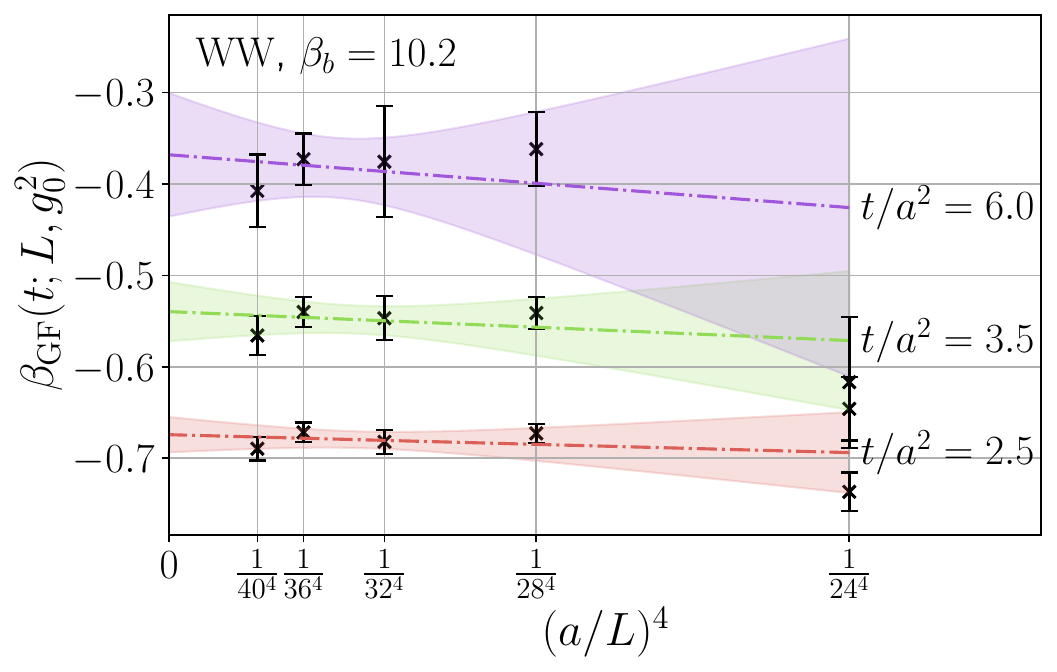}
    \caption{Result of our infinite volume extrapolation of $g^2_{\mathrm{GF}}\big(t;L,g_0^2\big)$ (left panels) and $\beta_{\mathrm{GF}}\big(t;L,g_0^2\big)$ (right panels) for the Wilson (W) operator at $\beta_{b}=9.60$ (top panels), $9.80$ (middle panels) and $10.2$ (bottom panels). Black ($\times$) markers with error bars are the data included in our extrapolation. Extrapolations with errors that are predicted from Bayesian model averaging are indicated by multi-colored bands at $t/a^2=2.5$ (red), $3.5$ (light green), $4.5$ (cyan), and $6.0$ (light purple). We do not show the infinite volume extrapolation of $\beta_{\mathrm{GF}}\big(t;L,g_0^2\big)$ at $t/a^2=4.5$ for visualization purposes.}
    \label{fig:inf_vol_W}
\end{figure*}

\begin{figure*}[tb]
    \centering
    \includegraphics[width=\columnwidth]{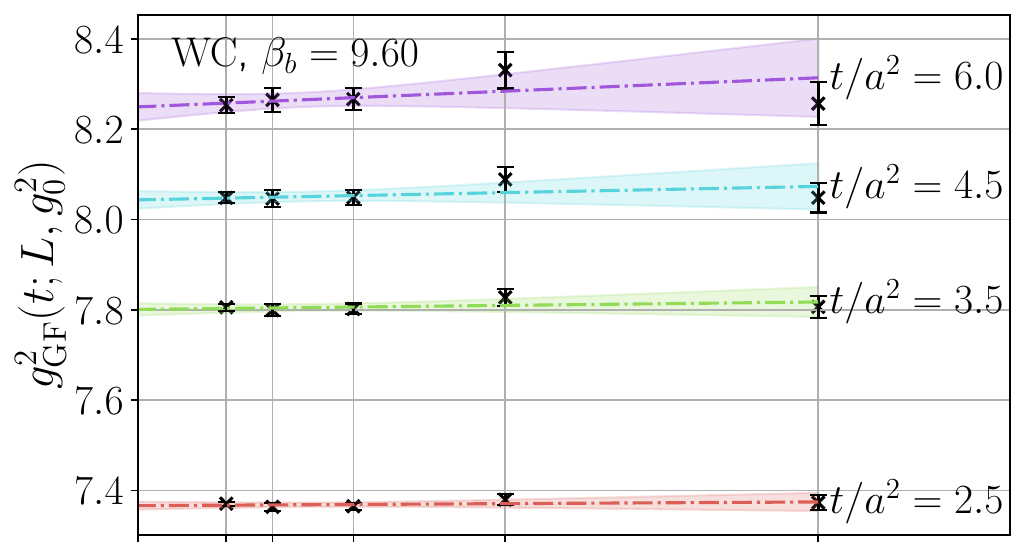}\includegraphics[width=\columnwidth]{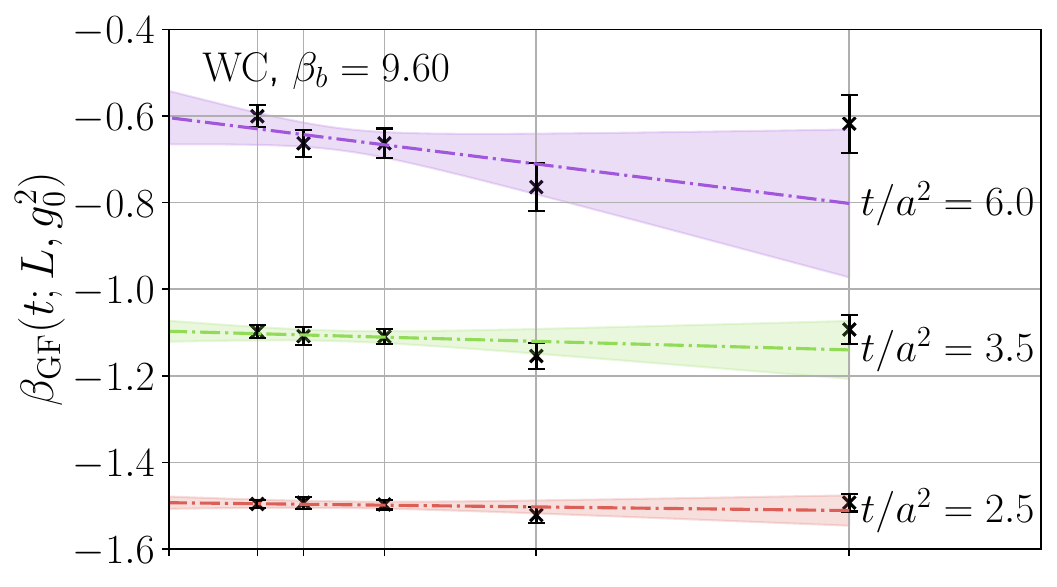}
    \includegraphics[width=\columnwidth]{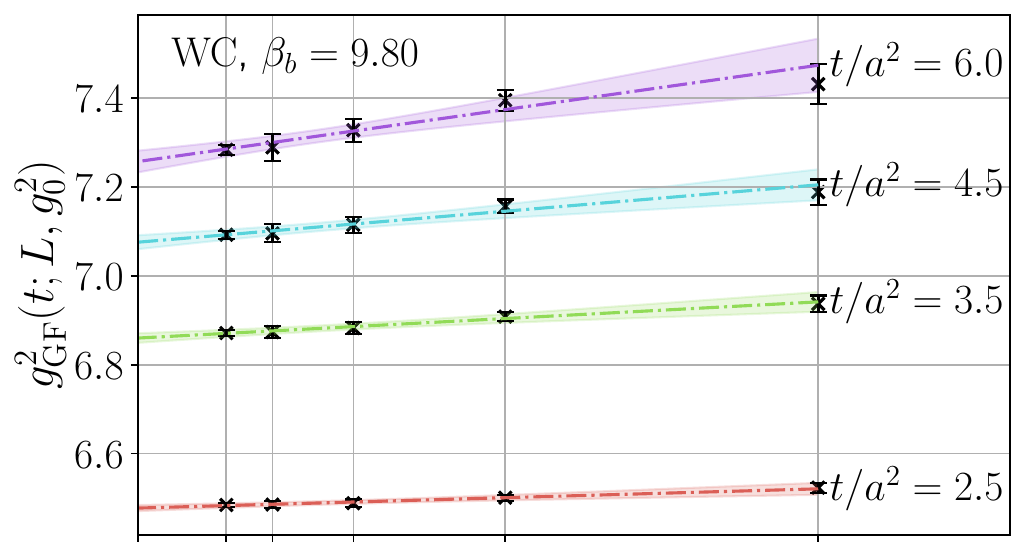}\includegraphics[width=\columnwidth]{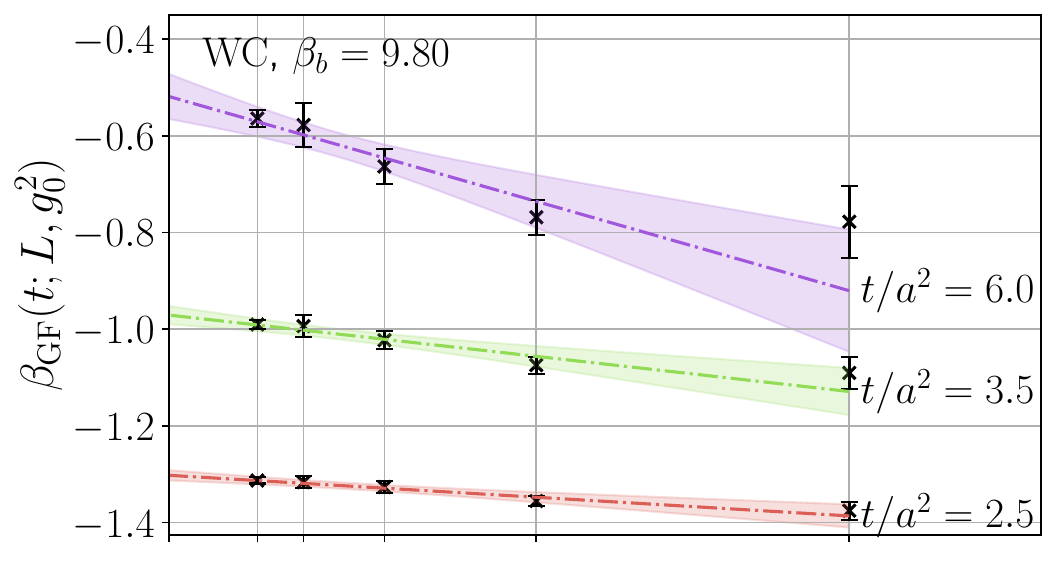}
    \includegraphics[width=\columnwidth]{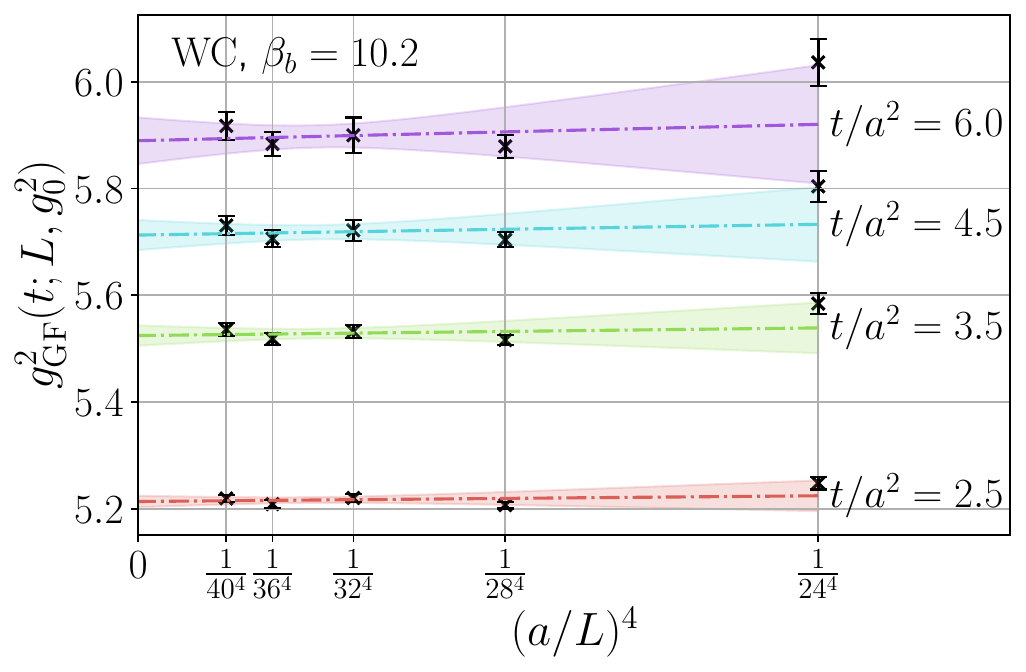}\includegraphics[width=\columnwidth]{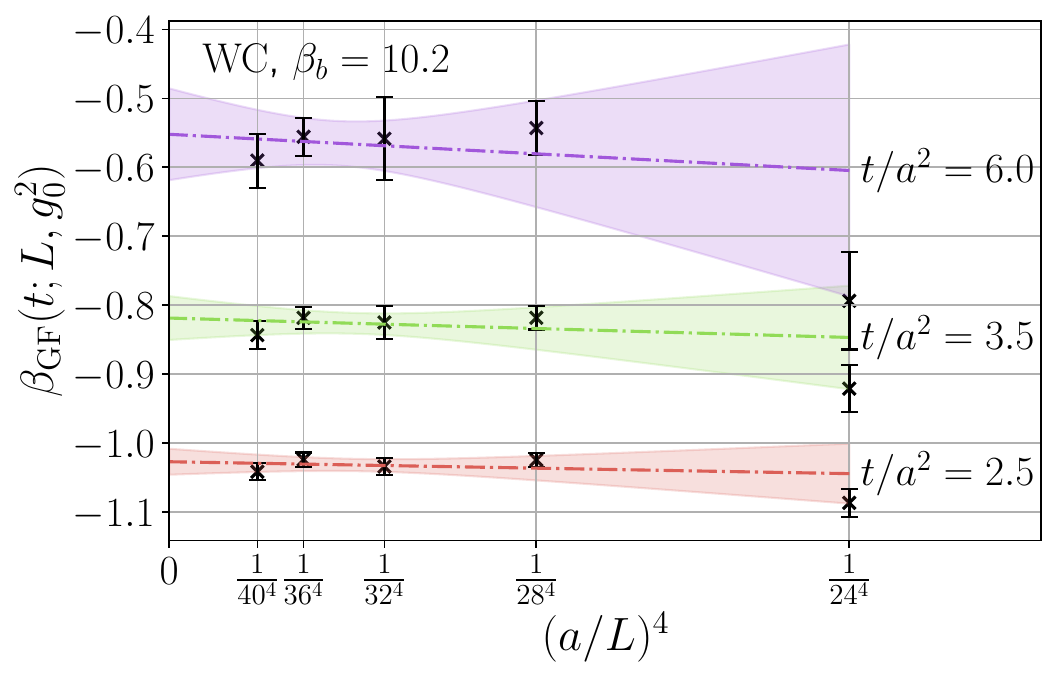}
    \caption{Result of our infinite volume extrapolation of $g^2_{\mathrm{GF}}\big(t;L,g_0^2\big)$ (left panels) and $\beta_{\mathrm{GF}}\big(t;L,g_0^2\big)$ (right panels) for the clover (C) operator at $\beta_{b}=9.60$ (top panels), $9.80$ (middle panels) and $10.2$ (bottom panels). Black ($\times$) markers with error bars are the data included in our extrapolation. Extrapolations with errors that are predicted from Bayesian model averaging are indicated by multi-colored bands at $t/a^2=2.5$ (red), $3.5$ (light green), $4.5$ (cyan), and $6.0$ (light purple). We do not show the infinite volume extrapolation of $\beta_{\mathrm{GF}}\big(t;L,g_0^2\big)$ at $t/a^2=4.5$ for visualization purposes.}
    \label{fig:inf_vol_C}
\end{figure*}

We measure the gradient flow coupling in finite volume in terms of the Yang-Mills energy density $E(t)$ as
\begin{equation}
    g^2_{\mathrm{GF}}\big(t;L,g_0^2\big) \equiv \frac{\mathcal{N}}{1+\delta(t,L)}\
\svev{t^2 E(t)},
\end{equation}
where $\delta(t,L)$ corrects for gauge zero modes \cite{Fodor:2012td}. From $g^2_{\mathrm{GF}}\big(t;L,g_0^2\big)$, we calculate the gradient flow $\beta$ function in finite volume as
\begin{equation}
     \beta_{\mathrm{GF}}\big(t;L,g_0^2\big) \equiv -t\frac{\mathrm{d}}{\mathrm{d}t} g^2_{\mathrm{GF}}\big(t;L,g_0^2\big),
\end{equation}
where we discretize $\mathrm{d}/\mathrm{d}t$ with a 5-point stencil.  The autocorrelation time for $g^2_{\mathrm{GF}}\big(t;L,g_0^2\big)$ and $\beta_{\mathrm{GF}}\big(t;L,g_0^2\big)$ is typically between 20-80 molecular dynamics time units (MDTUs), with occasional jumps to 120-200 MDTUs.

To extract the continuum $\beta_{\mathrm{GF}}\big(g^2_{\mathrm{GF}}\big)$ as a function of $g^2_{\mathrm{GF}}$, we follow the CBFM procedure outlined in Refs. \cite{Peterson:2021lvb, Hasenfratz:2023bok}.
\begin{enumerate}
    \item Take the infinite volume limit by independently extrapolating both $g^2_{\mathrm{GF}}\big(t;L,g_0^2\big)$ and $\beta_{\mathrm{GF}}\big(t;L,g_0^2\big)$ linearly in $(a/L)^4 \rightarrow 0$ at fixed $t/a^2$ and $\beta_b$.
    \item Interpolate $\beta_{\mathrm{GF}}\big(t;g_{0}^2\big)$ in $g^2_{\mathrm{GF}}\big(t;g_{0}^2\big)$ at fixed $t/a^2$.
    \item Take the continuum limit by extrapolating $\beta_{\mathrm{GF}}\big(t;g_{0}^2\big)$ linearly in $a^2/t \rightarrow 0$ at fixed $g^2_{\mathrm{GF}}(t)$.
\end{enumerate}
Correlated uncertainties are propagated throughout our analysis using the automatic error propagation tools provided by the \texttt{gvar} library \cite{gvar}. Fits are performed using the \texttt{SwissFit} library, which integrates directly with \texttt{gvar} \cite{Peterson_SwissFit}. The steps of the CBFM are detailed in the rest of this section.

\subsection{Infinite volume extrapolation}

Because the Yang-Mills energy density is a dimension-4 operator, leading finite-volume corrections to $g^2_{\mathrm{GF}}\big(t;L,g_0^2\big)$ and $\beta_{\mathrm{GF}}\big(t;L,g_0^2\big)$ are expected to be $\mathcal{O}\big(t^2/L^4\big)$. Therefore, we extrapolate both $g^2_{\mathrm{GF}}\big(t;L,g_0^2\big)$ and $\beta_{\mathrm{GF}}\big(t;L,g_0^2\big)$ to $a/L \rightarrow 0$ by independently fitting them to the ansatz 
\begin{equation}
    \mathrm{FV}(L/a)=k_1(a/L)^4+k_2
\end{equation}
at fixed $t/a^2$ and $\beta_b$. This analysis strategy was first outlined in Ref. \cite{Peterson:2021lvb} and subsequently applied in Refs. \cite{Wong:2023jvr, Hasenfratz:2023bok}. Alternative methods are discussed in Refs. \cite{Fodor:2017die, Hasenfratz:2019puu, Hasenfratz:2019hpg, Kuti:2022ldb}. 

We account for the systematic uncertainty that is associated with choosing a particular subset of volumes for the infinite volume extrapolation using Bayesian model averaging \cite{Jay:2020jkz, Neil:2022joj, Neil:2023pgt}. We do so by first fitting over all possible subsets of volumes $L/a \in \{24,28,32,36,40\}$ with at least three volumes in each subset $\eta$. We calculate the model weight $w_{\eta}$ for a particular subset $\eta$ as
\begin{equation}
    w_{\eta} \propto \exp\bigg[-\frac{1}{2}\big(\chi^2_{\eta}+2d_{\eta}\big)\bigg],
\end{equation}
where $\chi^2_\eta$ is the $\chi^2$ statistic of fit $\eta$ and $d_{\eta}$ is the number of data points \textit{not} included in fit $\eta$ from the full set of volumes $\{24,28,32,36,40\}$. Denoting the mean of $k_{i}$ from fit $\eta$ as $\overline{k}_{i}^{(\eta)}$, our model-averaged prediction for the mean $\overline{k}_i$ of $k_i$ is
\begin{equation}
    \overline{k}_i = \sum_{\eta} \overline{k}_i^{(\eta)} w_{\eta},
\end{equation}
where the weights $w_{\eta}$ have been normalized such that $\sum_{\eta} w_{\eta}=1$. The covariance $C_{ij}$ of our model-averaged prediction for $\{k_i\}_{i=1,2}$ is
\begin{equation}
    C_{ij}=\sum_{\eta}C_{ij}^{(\eta)}w_{\eta} + \sum_{\eta}\overline{k}_i^{(\eta)} \overline{k}_j^{(\eta)} w_{\eta}-\overline{k}_{i}\overline{k}_{j},
\end{equation}
where $C_{ij}^{(\eta)}$ is the covariance of $\{k_i^{(\eta)}\}_{i=1,2}$ from fit $\eta$.

In Figs. \ref{fig:inf_vol_W} and \ref{fig:inf_vol_C}, we show the result of our model-averaged infinite volume extrapolation for the W and C discretization of $E(t)$, respectively, over a range of flow times $2.5 \leq t/a^2 \leq 6.0$ (different colors). The left panels of Figs. \ref{fig:inf_vol_W}-\ref{fig:inf_vol_C} show our infinite extrapolation of $g^2_{\mathrm{GF}}\big(t;L,g_0^2\big)$, while the right panels show our infinite volume extrapolation of $\beta_{\mathrm{GF}}\big(t;L,g_0^2\big)$. The bare gauge couplings that we chose for these plots are in the vicinity where the continuum $\beta$ function predicts an IRFP. For all three bare gauge couplings shown in Figs. \ref{fig:inf_vol_W}-\ref{fig:inf_vol_C}, the model average is dominated by subsets containing $L/a \in \{28,32,36,40\}$, as $L/a=24$ often deviates from the linear trend in $a^4/L^4$, particularly as the flow time increases. This is reflected in the model average, as fits including $L/a=24$ possess small model weights $w_{\eta}$ and contribute negligibly to the model average. 

\subsection{Intermediate interpolation}\label{sec:intermediate_interpolation}

\begin{figure}[tb]
    \centering
    \includegraphics[width=\columnwidth]{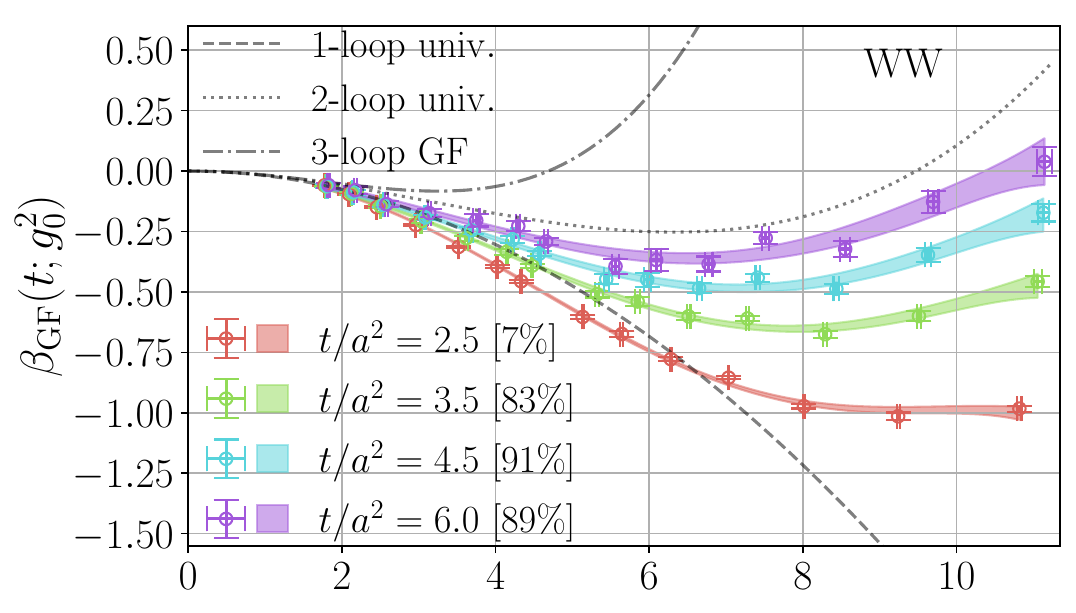}
    \includegraphics[width=\columnwidth]{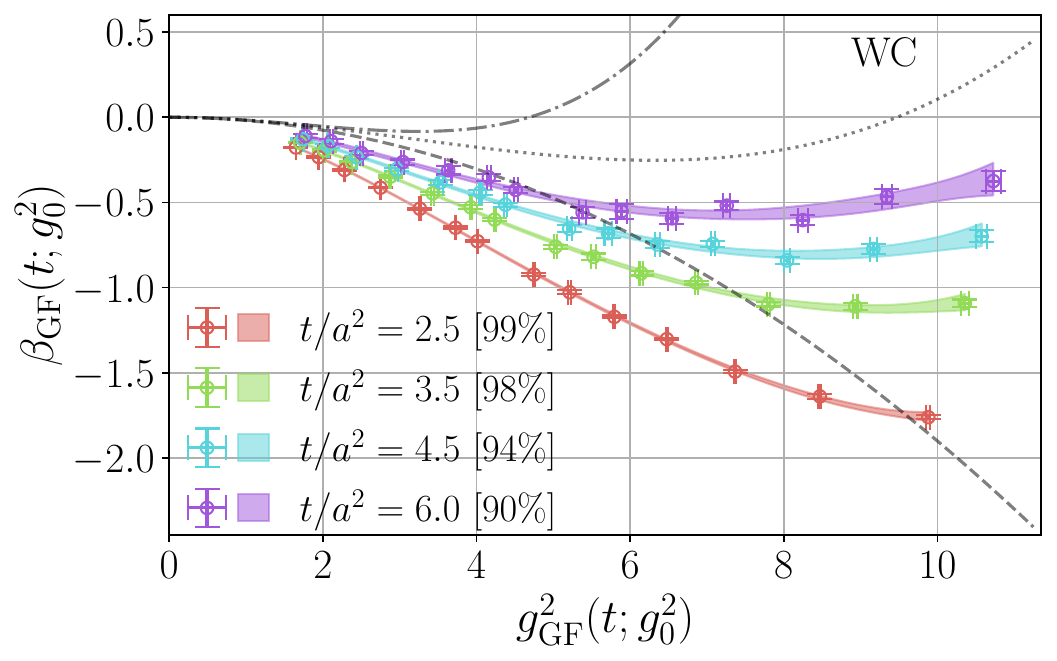}
    \caption{Illustration of our interpolation of $\beta_{\mathrm{GF}}\big(t;g_0^2\big)$ in $\beta_{\mathrm{GF}}\big(t;g_0^2\big)$ for the Wilson operator (top panel) and clover operator (bottom panel). Interpolations at fixed $t/a^2$ are indicated by colored bands, with $t/a^2=2.5$ (red), $3.5$ (light green), $4.5$ (cyan), and $6.0$ (light purple). The width of the band indicates the error. The data contributing to each interpolation is indicated by an open circular marker with both x- and y-errors. We compare our interpolation against the continuum 1- (dashed), 2- (dotted), and 3-loop (dashed-dotted) gradient flow $\beta$ function from perturbation theory \cite{Harlander:2016vzb}. }
    \label{fig:interp}
\end{figure}

The continuum limit $a^2/t \to 0$ of  $\beta_{\mathrm{GF}}\big(t;g_0^2\big)$ 
is taken at fixed $g^2_{\mathrm{GF}}$. We predict  pairs 
$\big(t/a^2,\beta_{\mathrm{GF}}\big(t;g_0^2\big)\big)$ at a set of fixed $g^2_{\mathrm{GF}}$ for the continuum extrapolation  by interpolating $\beta_{\mathrm{GF}}\big(t;g_0^2\big)$ in $g^2_{\mathrm{GF}}\big(t;g_0^2\big)$ at fixed $t/a^2$ using the ansatz
\begin{equation}
    \mathcal{I}_{N}\big(g^2_{\mathrm{GF}}\big) = g^4_{\mathrm{GF}}\sum_{i=0}^{N-1}p_{n}g^{2n}_{\mathrm{GF}}.
\end{equation}
At each $t/a^2$, we account for the uncertainty in $g^2_{\mathrm{GF}}\big(t;g_0^2\big)$ by including the mean and covariance of $g^2_{\mathrm{GF}}\big(t;g_0^2\big)$ as a Gaussian prior. We also set a Gaussian prior on each coefficient $p_{n}$ with zero mean and a width of $0.1$, which helps stabilize the fit. We choose $N=4$, as it is the lowest value of $N$ that fits the data well. In Fig. \ref{fig:interp}, we show the result of our interpolation for the W operator (top panel) and the C operator (bottom panel) for several flow time values in the range $2.5 \leq t/a^2 \leq 6.0$ (different colors). The fits that enter our continuum extrapolation have p-values in the $83\%-98\%$ range. This could indicate that we are either overfitting or the errors in our data are overestimated. Reducing the order makes each interpolation significantly worse, as interpolations with $N \leq 3$ are unable to accommodate the varying curvature at weak/strong coupling. Therefore, use $N=4$ for our central analysis. We will discuss the systematic effect that is associated with the order $N$ in our estimate of $g^2_{\mathrm{GF}\star}$ and $\gamma_{g}^{\star}$ in Sec. \ref{sec:leading_irrel_exp}.

\subsection{Continuum extrapolation}\label{eqn:cont_extrap}

\begin{figure}
    \centering
    \includegraphics[width=\columnwidth]{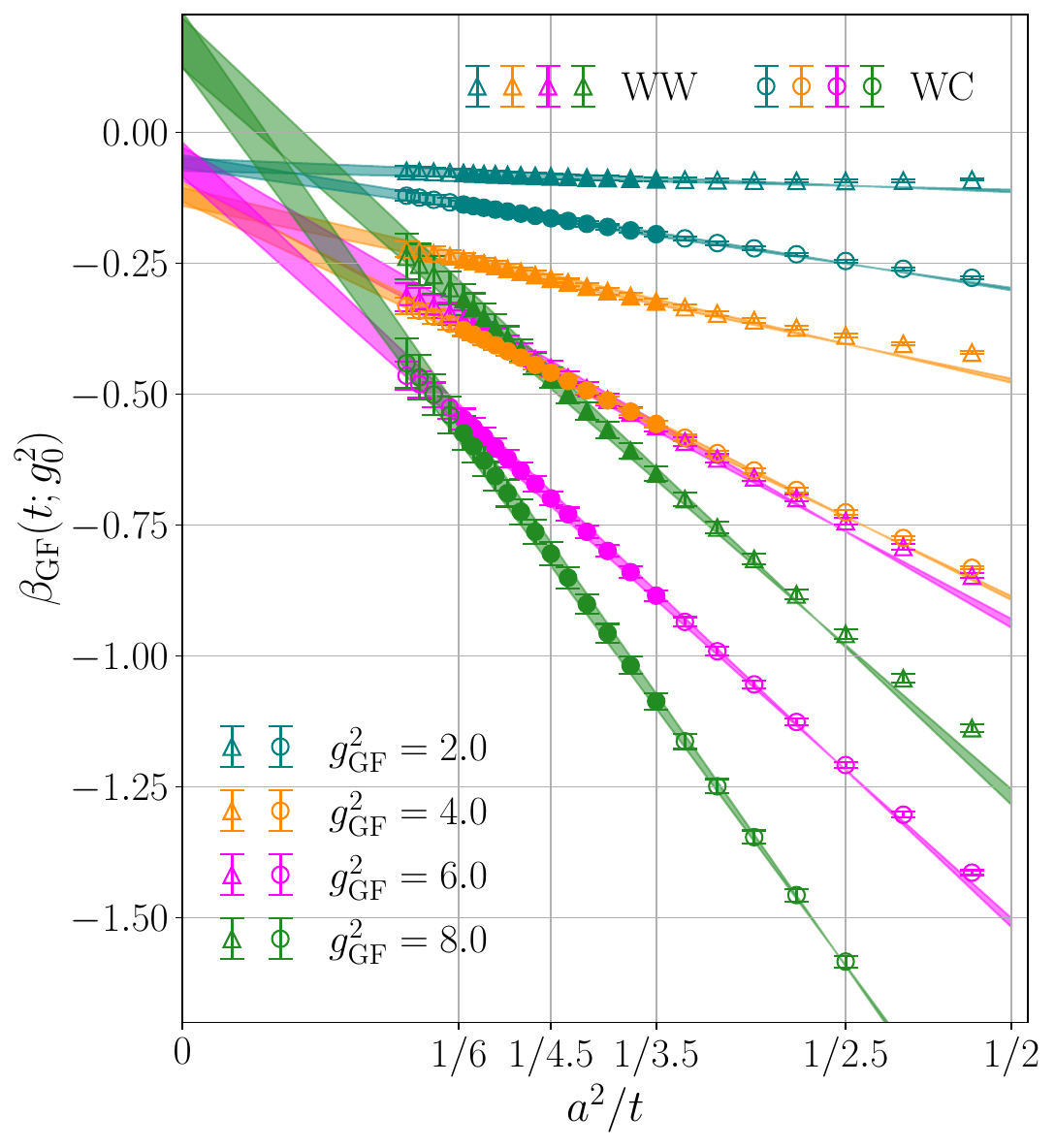}
    \caption{Illustration of our continuum extrapolation of $\beta_{\mathrm{GF}}\big(t;g_0^2\big)$ at fixed $g^2_{\mathrm{GF}}=2.0$ (teal), $4.0$ (dark orange), $6.0$ (magenta), and 8.0 (forest green). Data contributing to our extrapolation with the W operator are shown as error bars with triangular markers and the C operator are shown as error bars with circular markers. Our extrapolations are shown as colored bands, where the error is indicated by the width of the band.}
    \label{fig:explicit_cont_extrap}
\end{figure}

\begin{figure}
    \centering
    \includegraphics[width=\columnwidth]{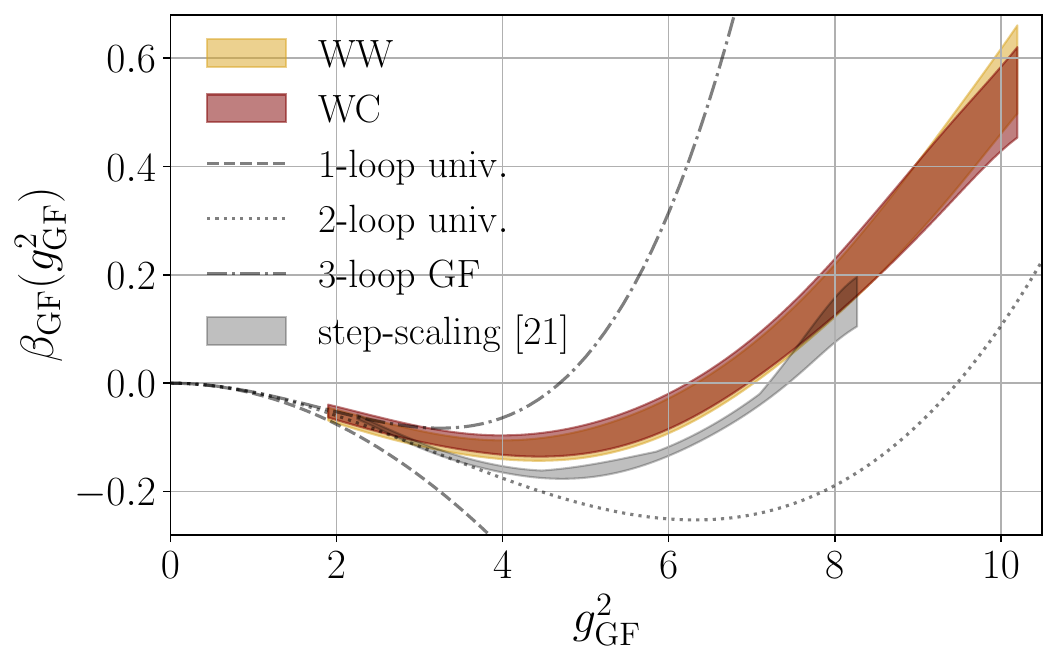}
    \caption{Our continuum prediction for $\beta_{\mathrm{GF}}\big(g^2_{\mathrm{GF}}\big)$ as a function of $g^2_{\mathrm{GF}}$ for the W operator (gold band) and C operator (maroon band). The width of the band indicates the error. The nonperturbative results are juxtaposed against the 1- (dashed), 2- (dotted), and 3-loop (dashed-dotted) gradient flow $\beta$ function from perturbation theory \cite{Harlander:2016vzb}. Also shown is the step-scaling $\beta$ function in the $c=0.25$ scheme from Ref. \cite{Hasenfratz:2016dou} as a grey band.}
    \label{fig:cont_extrap}
\end{figure}

The final step is the continuum ($a^2/t \rightarrow 0$) limit over a set of fixed $g^2_{\mathrm{GF}}$ that predicts $\beta_{\mathrm{GF}}(g^2_{\mathrm{GF}})$ as a function of $g^2_{\mathrm{GF}}$. The range of $[t_{\mathrm{min}}/a^2,t_{\mathrm{max}}/a^2]$ used in the continuum extrapolation must be chosen with care. The value of $t_{\mathrm{min}}/a^2$ must be large enough for the RG flow to reach the renormalized trajectory. Once this is the case, finite-cutoff effects are $\mathcal{O}(a^2/t)$. In practice, one can identify when $t_{\mathrm{min}}/a^2$ is close enough to the renormalized trajectory by the overlap between the continuum prediction for $\beta_{\mathrm{GF}}(g^2_{\mathrm{GF}})$ from both operators. Because finite-volume effects are expected to be $\mathcal{O}\big(t^2/L^4\big)$, $t_{\mathrm{max}}/a^2$ must be chosen such $t/a^2 < t_{\mathrm{max}}/a^2$ has a reliable infinite volume extrapolation. Ideally, we would apply Bayesian model averaging to the continuum extrapolation to automatically account for systematic effect that is associated with making a particular choice in $[t_{\mathrm{min}}/a^2,t_{\mathrm{max}}/a^2]$. However, at this point in the analysis, we no longer have access to the full covariance matrix, which means that we no longer have access to a reliable estimate of the model weights. To estimate the error in our continuum extrapolation, we use the half-difference of the prediction from the continuum extrapolation performed at $\pm 1\sigma$. This approach was also taken in Ref. \cite{Hasenfratz:2023bok}.

Fig. \ref{fig:explicit_cont_extrap} shows  examples of the continuum extrapolation performed in the range $2.0 \leq g^2_{\mathrm{GF}} \leq 6.0$ (different colors) using  $[t_{\mathrm{min}}/a^2,t_{\mathrm{max}}/a^2]=[3.5,6.0]$. For $t/a^2 \lesssim 3.5$, $\beta\big(t;g_0^2\big)$ has a slight curvature in $a^2/t$, indicating emerging higher-order cutoff effects for both the W and C operator. For $t/a^2 \gtrsim 6.0$, the data begins to deviate from a linear trend in $a^2/t$, indicating that the infinite volume extrapolation is getting unreliable. Our choice of $[t_{\mathrm{min}}/a^2,t_{\mathrm{max}}/a^2]=[3.5,6.0]$ avoids both of these two regimes. In Sec. \ref{sec:leading_irrel_exp}, we discuss the sensitivity of our prediction for $g^2_{\mathrm{GF}\star}$ and $\gamma_{g}^{\star}$ to our choice of $t_{\mathrm{min}}/a^2,t_{\mathrm{max}}/a^2$.

We show our prediction for the continuum $\beta_{\mathrm{GF}}\big(g^2_{\mathrm{GF}}\big)$ in Fig. \ref{fig:cont_extrap} from the WW (gold band) and WC (maroon band) combination. The continuum predictions from both operators are consistent with one another across the entire range of investigated renormalized couplings $g^2_{\mathrm{GF}}$. At small $g^2_{\mathrm{GF}}$, the continuum $\beta_{\mathrm{GF}}\big(g^2_{\mathrm{GF}}\big)$ appears to converge to the 1-, 2-, and 3-loop perturbative gradient flow $\beta$ function \cite{Harlander:2016vzb}. At $g^2_{\mathrm{GF}\star} \approx 6.60$, our continuum $\beta$ function predicts an infrared fixed point. The location of the fixed point is slightly below the predicted IRFP from the step-scaling calculation of Ref. \cite{Hasenfratz:2016dou}. Note that, because the calculation in Ref. \cite{Hasenfratz:2016dou} was done in a different gradient-flow-based renormalization scheme, the predicted $g^2_{\mathrm{GF}\star}$ values do not have to agree. Our final result for $\beta_{\mathrm{GF}}\big(g^2_{\mathrm{GF}}\big)$ from both operators is provided as an ASCII file. 

\section{The IRFP and its leading irrelevant critical exponent}\label{sec:leading_irrel_exp}

\begin{figure*}[tb]
    \centering
    \includegraphics[width=\columnwidth]{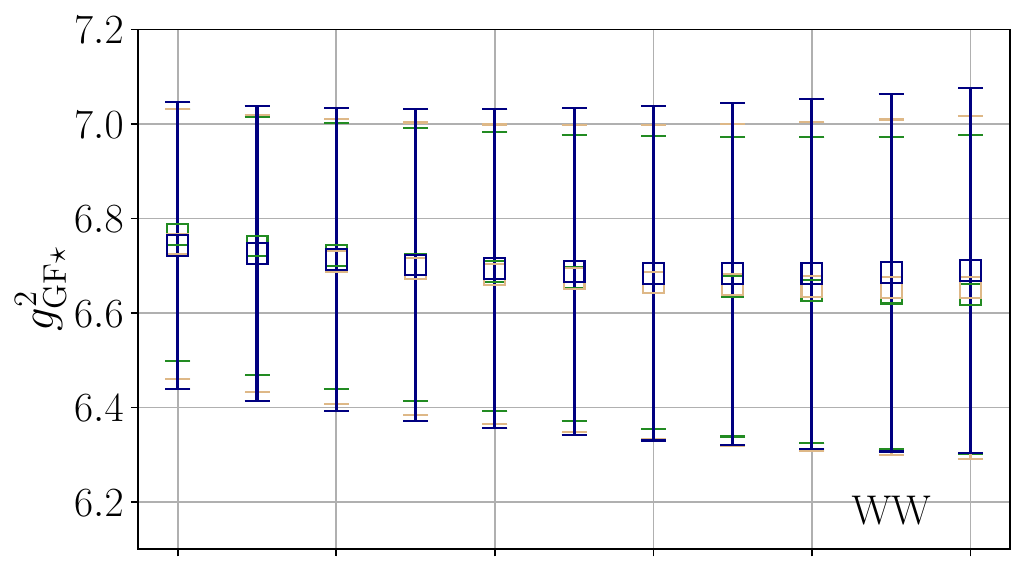}\includegraphics[width=0.89\columnwidth]{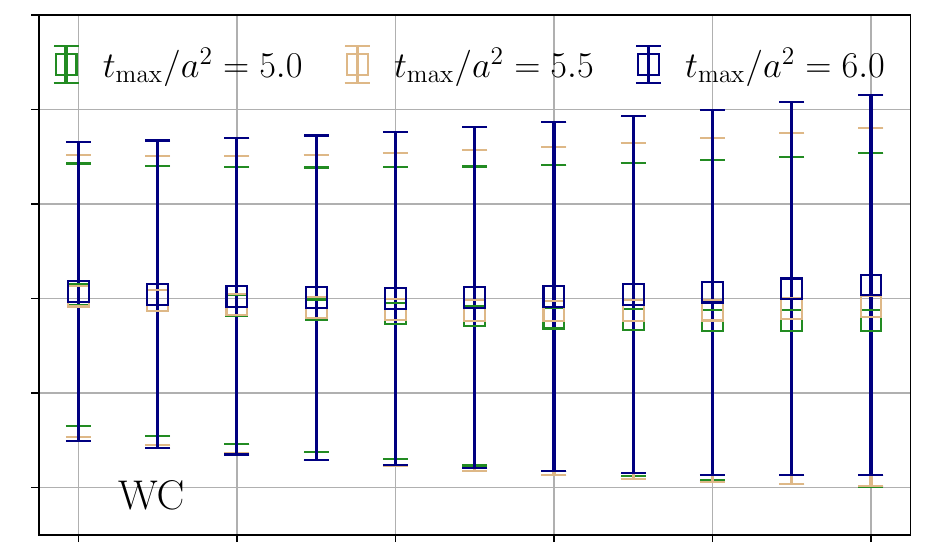}
    \includegraphics[width=\columnwidth]{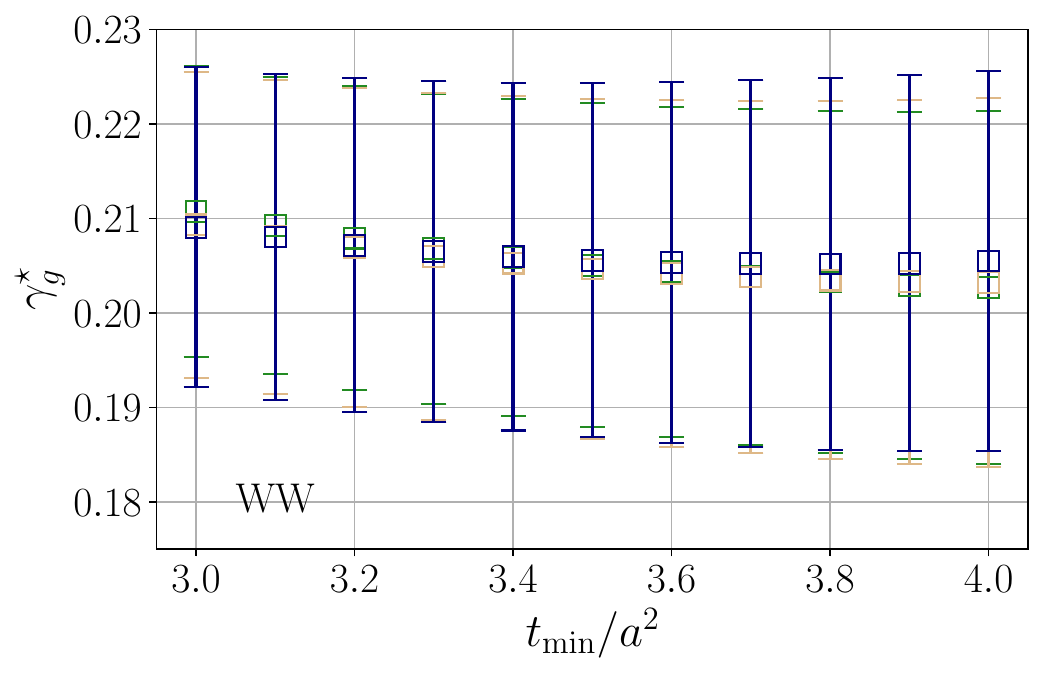}\includegraphics[width=0.88\columnwidth]{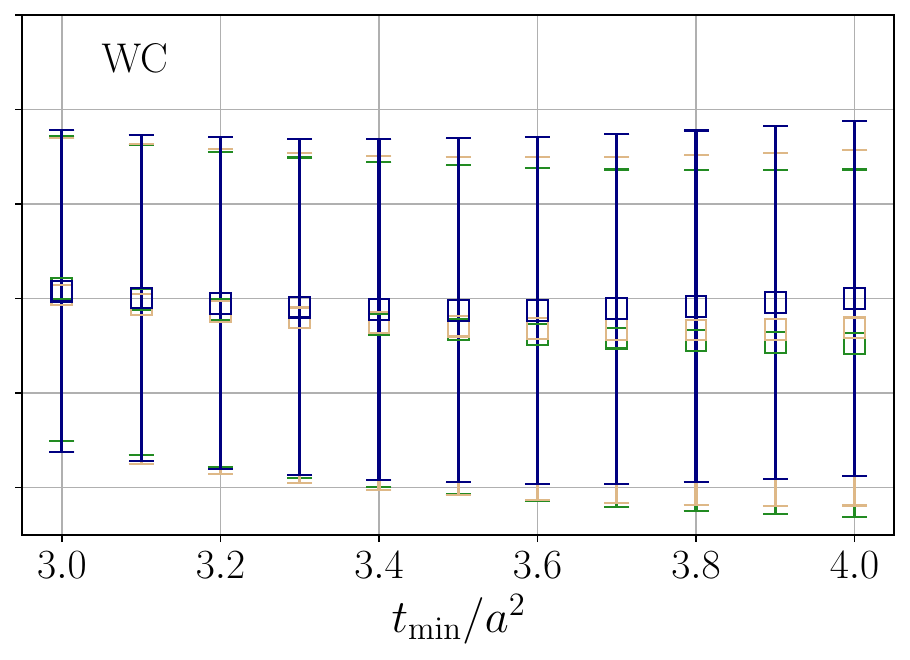}
    \caption{Comparison of our estimated $g^2_{\mathrm{GF}\star}$ and $\gamma_{g}^{\star}$ for different $t_{\mathrm{min}}/a^2$ (x-axes) and $t_{\mathrm{max}}/a^2=5.0$ (green), $5.5$ (gold), $6.0$ (navy) from the continuum extrapolation.}
    \label{fig:tmin_tmax_variation}
\end{figure*}

\begin{figure}
    \centering
    \includegraphics[width=\columnwidth]{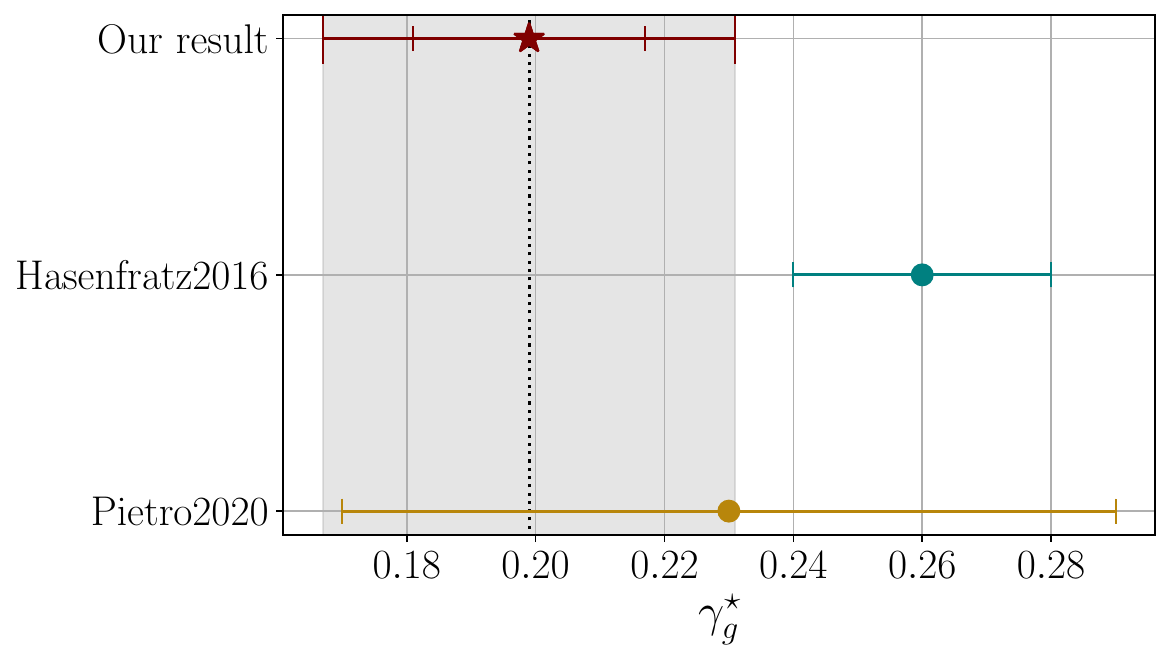}
    \caption{Comparison of our value for $\gamma_{g}^{\star}$ (maroon errorbar) against Ref. \cite{Hasenfratz:2016dou} (teal error bar) and Ref. \cite{DiPietro:2020jne} (dark gold error bar). The smaller error bar on our result indicates the error without accounting for systematic effects; the larger error bar indicates our error after accounting for systematic effects. We indicate our total error with a grey band for visualization.}
    \label{fig:lit_comparison}
\end{figure}

In the vicinity of the RG fixed point $g^2_{\mathrm{GF}\star}$
\begin{equation}
    \beta\big(g^2_{\mathrm{GF}}\big) \approx \frac{\gamma_{g}^{\star}}{2}\big(g^2_{\mathrm{GF}}-g^2_{\mathrm{GF}\star}\big), \quad g^2_{\mathrm{GF}} \to g^2_{\mathrm{GF}\star}
\end{equation}
where $\gamma_{g}^{\star}$ is the universal critical exponent of the irrelevant gauge coupling. The factor of $1/2$ is chosen to match the convention of Refs. \cite{Hasenfratz:2016dou, DiPietro:2020jne}. We estimate $g^2_{\mathrm{GF}\star}$ and $\gamma_{g}^{\star}$ via the following procedure.
\begin{enumerate}
    \item Interpolate the central value of $\beta\big(g^2_{\mathrm{GF}}\big)$ and the central value of $\beta\big(g^2_{\mathrm{GF}}\big) \pm 1\sigma$ in $g^2_{\mathrm{GF}}$ using a monotonic spline.
    \item Estimate the central value of $g^2_{\mathrm{GF}\star}$ from the root of the spline interpolation of $\beta\big(g^2_{\mathrm{GF}}\big)$ in $g^2_{\mathrm{GF}}$.
    \item Estimate the central value of $\gamma_{g}^{\star}$ from the derivative of the spline at $g^2_{\mathrm{GF}\star}$.
    \item Repeat Steps (2)-(3) with a spline interpolation of $\beta\big(g^2_{\mathrm{GF}}\big) \pm 1\sigma$ in $g^2_{\mathrm{GF}}$. 
    \item Estimate the error in $g^2_{\mathrm{GF}\star}$ and $\gamma_{g}^{\star}$ from the half difference of their predictions from the interpolations in Step (4).
\end{enumerate}
Steps (1)-(5) yield $g^2_{\mathrm{GF}\star}=6.69(35),6.60(36)$ and $\gamma_{g}^{\star}=0.206(19),0.199(18)$ for WW and WC, respectively. In Fig. \ref{fig:tmin_tmax_variation} we look at how the continuum limit predictions for $g^2_{\mathrm{GF}\star}$ (top panels) and $\gamma_{g}^{\star}$ (bottom panels)  vary with our choice of $t_{\mathrm{min}}/a^2$ (x-axes) and $t_{\mathrm{max}.}/a^2$ (different colors) for the W (left panels) and C (right panels) operators. The central values for both quantities and both operators are stable; they vary well within error. The stability in $t_{\mathrm{min}}/a^2$ is attributed to the linearity of the continuum extrapolation over a wide range of $a^2/t$, while the stability in $t_{\mathrm{max}}/a^2$ is likely attributed to our control over the infinite volume extrapolation. 

We take the result for $\gamma_{g}^{\star}$ from the WC combination with the value for $[t_{\mathrm{min}}/a^2,t_{\mathrm{max}}/a^2]=[3.5,6.0]$ from Sec. \ref{sec:conclusions} as our central result. We estimate additional systematic errors by varying our analysis as follows.
\begin{itemize}
    \item Choosing a higher-order polynomial for the intermediate interpolation in Sec. \ref{sec:intermediate_interpolation}. The highest-order interpolation in $N$ that we can use before we lose control over our continuum extrapolation due to overfitting is $N=6$. This shifts the value of $\gamma_{g}^{\star}$ by $\approx 0.001$, and we take the latter difference as an estimate for the systematic error that is associated with our choice of $N$ for the intermediate interpolation.
    \item Choosing a different $t_{\mathrm{min}}/a^2,t_{\mathrm{max}}/a^2$ in the continuum extrapolation. We estimate the systematic error that is associated with our choice of $t_{\mathrm{min}}/a^2,t_{\mathrm{max}}/a^2$ by the difference in the most extreme values of $\gamma_{g}^{\star}$ in our variations illustrated in Fig. \ref{fig:tmin_tmax_variation}. This yields a systematic error of $\approx 0.006$.
    \item Choosing instead the prediction for $\gamma_{g}^{\star}=0.206(19)$ from the WW combination as our central result. We take the difference in these predictions ($\approx 0.007$) as an estimate of the systematic error associated with making a particular choice of flow/operator combination.
\end{itemize}
To be conservative, we combine the error in our analysis of $\gamma_{g}^{\star}$ with the systematic error estimates above linearly. This yields the final of prediction $\gamma_{g}^{\star}=0.199(32)$. Repeating the same exercise for $g^2_{\mathrm{GF}\star}$ yields a systematic error of $\approx 0.12$ from the interpolation order, $\approx 0.05$ from the continuum extrapolation, and $\approx 0.09$ from the flow/operator combination. Including the systematic error in $g^2_{\mathrm{GF}\star}$ linearly yields a final prediction of $g^2_{\mathrm{GF}\star}=6.60(62)$.

In Fig. \ref{fig:lit_comparison} we compare our prediction for $\gamma_{g}^{\star}$ with those available in the literature \cite{Hasenfratz:2016dou,DiPietro:2020jne}. 
Our result is plotted as a maroon star with errors indicated by an error bar. The smaller error bar is our error estimate before accounting for systematic effects and the larger error bar includes systematic effects. The result for $\gamma_{g}^{\star}$ from the perturbative calculation of Ref. \cite{DiPietro:2020jne} (dark gold error bar) is within $1\sigma$ of our estimate for $\gamma_{g}^{*}$. 
The lattice calculation of $\gamma_{g}^{\star}$ from Ref. \cite{Hasenfratz:2016dou} (cyan error bar) is within $2\sigma$ of our result. However, it is important to note that the lattice calculation in Ref. \cite{Hasenfratz:2016dou} uses smaller volumes and very coarse lattices in comparison to the present work. It is also worth noting the ``scheme-independent'' prediction of $\gamma_{g}^{\star} \approx 0.228$ from Ref. \cite{Ryttov:2017kmx} is also within $1\sigma$ of our predicted $\gamma_{g}^{\star}$; we don't show this result in Fig. \ref{fig:lit_comparison} because no estimate of the systematic error in this result is available.

\section{Conclusions}\label{sec:conclusions}

We have calculated the non-perturbative $\beta$ function of the SU(3) gauge-fermion system with twelve massless fundamental fermions using a Pauli-Villars improved lattice action. We find strong evidence for an infrared fixed point at $g^2_{\mathrm{GF}\star}=6.60(62)$ from our gradient-flow-based renormalization scheme. Our study utilizes a wide range of couplings and volumes. In particular, we can reach renormalized gauge coupling values well above the predicted IRFP without the interference of a bulk phase transition. We include systematic effects from the infinite volume extrapolation directly into our analysis using Bayesian model averaging. 
Our data exhibits cutoff effects that are consistent with the leading $\mathcal{O}(a^2/t)$ form over a wide range $t/a^2$. The consistency between the W and C operators further supports the leading order scaling behavior. We believe the improved scaling is due to the additional PV bosons that reduce cutoff effects. In contrast, we found significantly larger cutoff effects when we re-analyzed the data that were generated without PV fields and used in Ref. \cite{Hasenfratz:2016dou}. Our data is publicly available at Ref. \cite{peterson_2024_10719052}.

Overall, the systematics of our continuum extrapolation are well controlled. Based on the continuum prediction for $\beta\big(g^2_{\mathrm{GF}}\big)$ in $g^2_{\mathrm{GF}}$, we estimate the leading irrelevant critical exponent $\gamma_{g}^{\star}=0.199(32)$. This estimate includes conservative systematic errors from various choices in our analysis. Our result for $\gamma_{g}^{\star}$ agrees with Refs. \cite{Ryttov:2017kmx, DiPietro:2020jne} at the $1\sigma$ level and Ref. \cite{Hasenfratz:2016dou} at the $2\sigma$ level.

\begin{acknowledgments}

 Both authors acknowledge support by DOE Grant No.~DE-SC0010005. This material is based upon work supported by the National Science Foundation Graduate Research Fellowship Program under Grant No.~DGE 2040434. The research reported in this work made use of computing and long-term storage facilities of the USQCD Collaboration, which are funded by the Office of Science of the U.S. Department of Energy.  This work utilized the Alpine high-performance computing resource at the University of Colorado Boulder. Alpine is jointly funded by the University of Colorado Boulder, the University of Colorado Anschutz, and Colorado State University. We thank James Osborn and Xiaoyong Jin for writing \texttt{QEX} and helping us develop our \texttt{QEX}-based hybrid Monte Carlo and gradient flow codes.

\end{acknowledgments}

\bibliography{./BSM}

\end{document}